\documentclass[runningheads,a4paper]{llncs}

\usepackage{amsmath}
\usepackage{fixltx2e}
\usepackage{verbatim}

\usepackage{color}

\usepackage{subfigure}

\usepackage{amssymb}
\usepackage{graphicx}
\usepackage[font={normalsize},labelfont=bf]{caption}
\usepackage{varwidth}
\usepackage[hyphens]{url}
\makeatletter
\g@addto@macro{\UrlBreaks}{\UrlOrds}
\makeatother
\PassOptionsToPackage{hyphens}{url}
\usepackage{fancyvrb}
\fvset{frame=single,samepage=true,fontfamily=courier,numbers=left,numbersep=2pt,baselinestretch=0.50,fontsize=\small}

\urldef{\mailsa}\path|{maturban, mln, mweigle}@cs.odu.edu|
\newcommand{\keywords}[1]{\par\addvspace\baselineskip
\noindent\keywordname\enspace\ignorespaces#1}

\newcommand\userinput[1]{\textbf{#1}}

\usepackage{lipsum}

\newcommand\blfootnote[1]{%
  \begingroup
  \renewcommand\thefootnote{}\footnote{#1}%
  \addtocounter{footnote}{-1}%
  \endgroup
}

\begin{document}

\mainmatter  

\title{Quantifying Orphaned Annotations in Hypothes.is}

\titlerunning{Quantifying Orphaned Annotations in Hypothes.is}

%
%
\author{Mohamed Aturban \and Michael L. Nelson \and Michele C. Weigle}
\authorrunning{Mohamed Aturban \and Michael L. Nelson \and Michele C. Weigle}

\institute{Dept of Computer Science, Old Dominion University, Norfolk, VA, 23529
\mailsa\\
\url{http://ws-dl.cs.odu.edu}}

%
%

\toctitle{Quantifying Orphaned Annotations in Hypothes.is}
\tocauthor{Quantifying Orphaned Annotations in Hypothes.is}
\maketitle

\setcounter{secnumdepth}{3}

\begin{abstract}
Web annotation has been receiving increased attention recently with the organization of the Open Annotation Collaboration and new tools for open annotation, such as Hypothes.is. We investigate the prevalence of \emph{orphaned annotations}, where neither the live Web page nor an archived copy of the Web page contains the text that had previously been annotated in the Hypothes.is annotation system (containing 20,953 highlighted text annotations).  We found that about 22\% of highlighted text annotations can no longer be attached to their live Web pages.  Unfortunately, only about 12\% of these annotations can be reattached using the holdings of current public web archives, leaving the remaining 88\% of these annotations orphaned. For those annotations that are still attached, 53\% are in danger of becoming orphans if the live Web page changes.  This points to the need for archiving the target of an annotation at the time the annotation is created.
\end{abstract}

\section{Introduction}
\blfootnote{This arXiv paper is an extended version of our TPDL 2015 paper \cite{aturbanquantifying2015}.}
Annotating web resources helps users share, discuss, and review information and exchange thoughts. Haslhofer et al. \cite{haslhofer2011open} define annotation as associating extra pieces of information with existing web resources while the Open Annotation Collaboration (OAC) group defines an annotation as a set of connected resources \cite{haslhofer2011open} where the basic form of annotation consists of the Body and Target resources. Ideally, the Body should be about the Target. Annotation types include commenting on a web resource, highlighting text, replying to others' annotations, specifying a segment of interest rather than referring to the whole resource, tagging, etc.

Hypothes.is\footnote{http://hypothes.is}, an open annotation tool, was released in early 2013 and is publicly accessible for users to annotate, discuss, and share information. It provides different ways to annotate a web resource: highlighting text, adding notes, and commenting on and tagging a web page. In addition, it also allows users to share an individual annotation URI with each other as an independent web resource. The annotation is provided in JSON format and includes the annotation author, creation date, target URI, annotation text, permissions, tags, comments, etc.

One of the well-known issues of the Web is that Web pages are not fixed resources. A year after publication, about 11\% of content shared on social media will be gone \cite{salaheldeen2012losing,tpdl13:revolution}, and 20\% of scholarly articles have some form of reference rot \cite{klein2014scholarly}. Lost or modified web pages may result in \emph{orphaned annotations}, which can no longer be attached to their target web pages. 

\begin{figure}[h!]
\centering
\includegraphics[width=100mm]{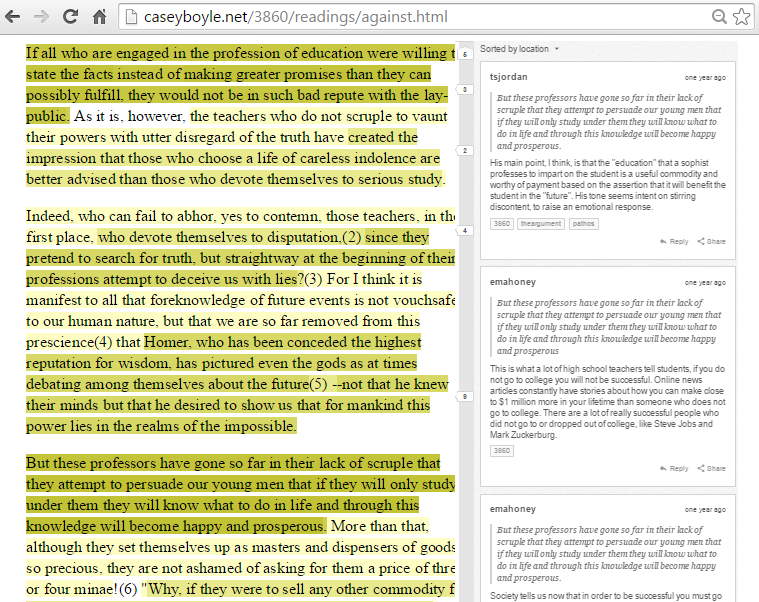}
\caption{Using the Hypothes.is Browser Extension to View the Annotations of http://caseyboyle.net/3860/readings/against.html}
\label{img:ann144}
\end{figure}

\begin{figure}[h!]
\centering
\includegraphics[width=122mm]{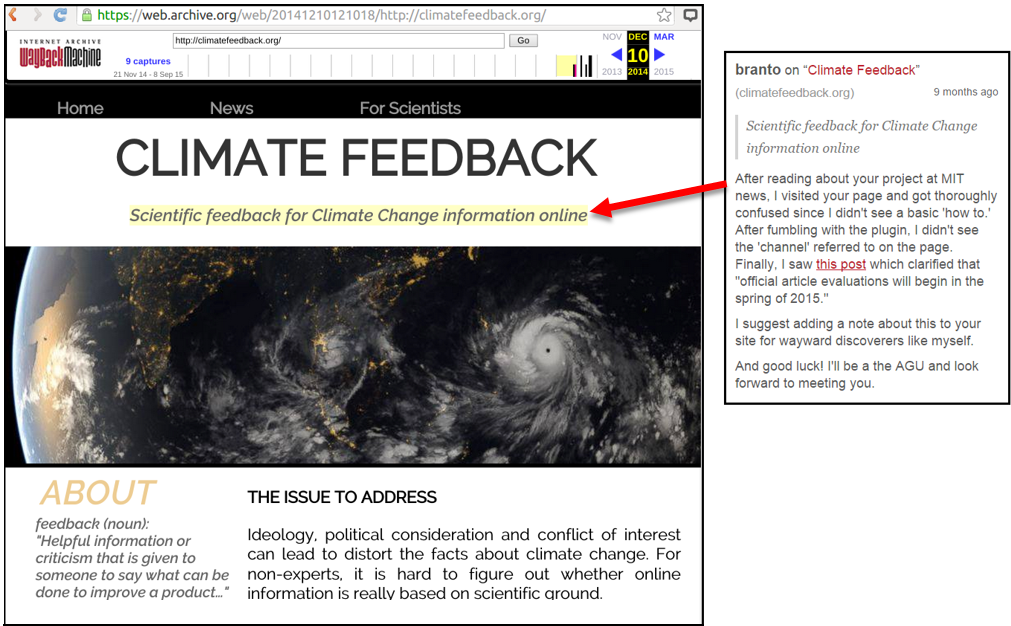}
\caption{http://climatefeedback.org/ in December 2014}
\label{img:arch-page}
\end{figure}

\begin{figure}[h!]
\centering
\includegraphics[width=122mm]{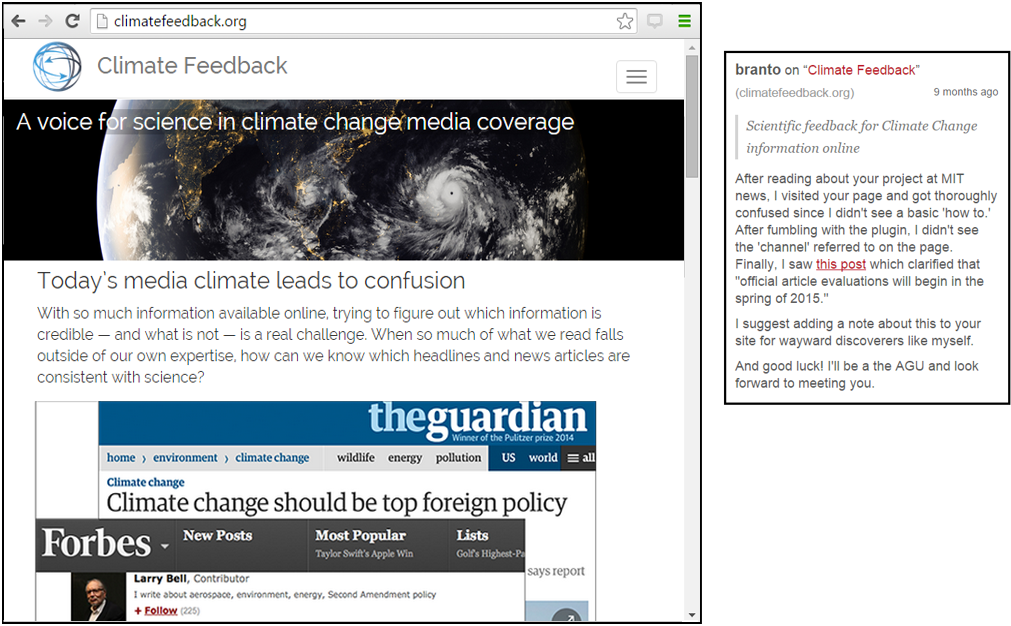}
\caption{http://climatefeedback.org/ in August 2015}
\label{img:live-page}
\end{figure}

Figure \ref{img:ann144} shows the annotated web page \url{http://caseyboyle.net/3860/readings/against.html} which has 144 annotations from Hypothes.is. The text with darker highlights indicates more users have selected this part of the page to annotate. The issue here is that all of these annotations are in danger of being orphaned because no copies of the target URI are available in the archives. Figure \ref{img:arch-page} shows the target URI \url{http://climatefeedback.org/}, created in December 2014, with the annotation ``After reading about your project at MIT news, I visited your page and ...'' on the highlighted text ``Scientific feedback for Climate Change information online''. In August 2015, this annotation can no longer be attached to the target web page because the highlighted text no longer appears on the page, as shown in Figure \ref{img:live-page}. Although the live Web version of \url{http://climatefeedback.org/} has changed and the annotation was in danger of being orphaned, the original version that was annotated has been archived and is available at the Internet Archive (\url{https://web.archive.org/web/20141210121018/http://climatefeedback.org/}). The annotation could be re-attached to this archived resource.

\begin{figure}[h!]
\centering
\includegraphics[width=90mm]{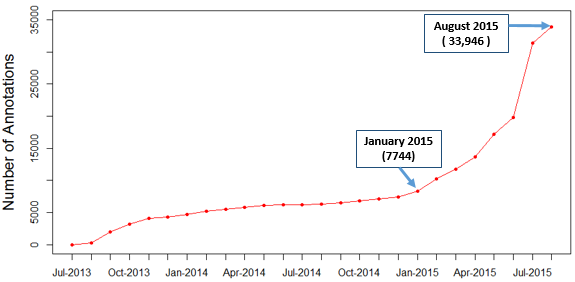}
\caption{Number of Annotations in Hypothes.Is since July 2013}
\label{img:tim}
\end{figure}

This paper is a follow-up to our paper \cite{aturbanquantifying2015} presented in August 2015 in the 19th International Conference on Theory and Practice of Digital Libraries (TPDL). In the TPDL version, we analyzed 6281 highlighted text annotations collected in January 2015, while in this paper, we worked with 20,953 annotations collected in August 2015. Figure \ref{img:tim} shows that the number of annotations in Hypothes.is has increased since July 2013. In this paper, we introduce a detailed analysis of the extent of orphaned highlighted text annotations in the Hypothes.is annotation system as of August 2015. We also look at the potential for web archives to be used to reattach these annotations. We find that 22\% of the highlighted text annotations at Hypothes.is are not attached to the live web, and only a few can be reattached using web archives.  Further, we show that 53\% of the currently attached annotations could potentially become orphans if their live Web resources change, because there are no archived versions of the annotated resources available.  Our analysis points to the potential for reducing orphaned annotations by archiving web resources at the time of annotation.

\section{Related Work}
Annotation has long been recognized as an important and fundamental aspect of hypertext systems \cite{marshall1998toward} and an integral part of digital libraries \cite{agosti2004annotations}, but broad adoption of general annotation for the Web has been slow. Annotations have been studied for digital library performance \cite{frommholz2006probabilistic,soo2003automated} and methods have been explored for aligning annotations in modified documents \cite{brush2001robust}, but typically such studies are limited to annotation systems specific for a particular digital library. While orphaned annotations of general web pages have been studied in the context of Walden's Paths \cite{furuta1997hypertext,francisco2001managing}, our study of Hypothes.is is a more recent evaluation of annotation and page synchronization in a widely deployed system.

Memento \cite{nelson:memento:tr} is an HTTP protocol extension that aggregates information about the resources available in multiple Web archives.  We can use Memento to obtain a list of archived versions of resources, or mementos, available in several web archives.  In this paper, we use the following Memento terminology:
\begin{itemize}
\item {URI-R - the original resource as it used to appear on the live Web.  A URI-R may have 0 or more mementos (URI-Ms)}
\item {URI-M - an archived snapshot of the URI-R at a specific date and time, which is called the Memento-Datetime, \emph{e.g.}, \emph{URI-M}$_i=$ \emph{URI-R}$@t_i$}
\item {TimeMap - a resource that provides a list of mementos (URI-Ms) for a URI-R, ordered by their Memento-Datetimes}
\end{itemize} 

There has been previous work in developing annotation systems to support collaborative work among users and in integrating the Open Annotation Data Model \cite{sanderson2013designing} with the Memento framework. The Open Annotation Collaboration (OAC) \cite{haslhofer2011open} has been introduced to make annotations reusable through different systems like Hypothes.is. Before publishing OAC, annotations would not be useful if the annotated web pages were lost because annotations were not assigned URIs independent from the web pages' URIs. By considering annotations as first-class web resources with unique URIs, annotation not only would become reusable if their targets disappear, but also would support interactivity between systems. Sanderson and Van de Sompel \cite{sanderson2010making} built annotation systems that support making web annotations persistent over time. They focus on integrating features in the Open Annotation Data Model with the Memento framework to help reconstructing annotations for a given memento and retrieving mementos for a given annotation. They did not focus on the case of orphaned annotations and assumed that the archived resources were available in web archives. Ainsworth et al. have estimated how much of the web is archived \cite{ainsworth2011much}. The result indicated that 35-90\% of publicly accessible URIs have at least one archived copy, although they did not consider annotations in their work, the result might estimate the number of orphaned annotations by factors like how frequently web pages are archived and the archiving process coverage. In other work \cite{sanderson2011sharedcanvas,haslhofer2014open} researchers built annotation systems that can deliver a better user experience for specialized users and scholars. The interfaces allow users to annotate multimedia web resources as well as medieval manuscripts in a collaborative way. In this paper, we focus on orphaned annotations and investigate how web archives could be used to reattach these annotations to the original text.

\section{Methodology}
We performed our analysis on the publicly accessible annotations available at Hypothes.is as of August 2015. The interface allows users to create different types of annotations: (1) making a note by highlighting text and then adding comments and tags about the selected text, (2) creating highlights only, (3) adding comments and tags without highlighting text, and (4) replying to an existing annotation. Table \ref{tab:annotation-types} shows how many annotations belong to each type.

\begin{table}
\centering
\renewcommand{\arraystretch}{1}
\begin{tabular}{r @{\hskip 0.2cm} |  @{\hskip 0.2cm} c @{\hskip 0.5cm} c @{\hskip 0.5cm} c}
\hline
\textbf{\shortstack{Number of\\Annotations}} & \textbf{\shortstack{Highlighted Text}} & \textbf{\shortstack{ Notes}} & \textbf{\shortstack{Tags}}\\
\hline
11,289 &  & \checkmark & \\
9858 & \checkmark & \checkmark & \\
9252 & \checkmark & \checkmark & \checkmark\\
1835 & \checkmark &  & \checkmark\\
1356 &  & \checkmark & \checkmark\\
348 &  &  & \checkmark\\
8 & \checkmark &  & \\
\hline
Total (33,946) & & & \\
\hline
\end{tabular}
\vspace{0.5em}
\caption {Annotation Types in Hypothes.is}
\label{tab:annotation-types}
\end{table}

In August 2015, we downloaded the JSON of all 33,946 publicly available annotations from Hypothes.is. Figure \ref{img:json} shows the JSON of the annotation from Figure \ref{img:arch-page} with relevant fields shown in bold.  The ``\texttt{updated}'' field gives the annotation creation date, ``\texttt{source}'' provides the annotation target URI, ``\texttt{type}''\texttt{:}``\texttt{TextQuoteSelector}'' indicates that it is a highlighted text annotation, ``\texttt{exact}'' contains the highlighted text, and ``\texttt{text}'' contains the annotation text itself. We focus only on annotations with highlighted text (``\texttt{type}''\texttt{:}``\texttt{TextQuoteS-\\elector}''), leaving 20,953 annotations for analysis. To determine how many of those annotations are orphaned, for each annotation we performed the following steps:
\begin{itemize}
\item Determine the current HTTP status of the annotation target URIs (``\texttt{source}'').
\item Compare selected highlighted text (``\texttt{exact}'') to the text of the current version of the URI.
\item Discover available mementos for the target URI.
\item Search for highlighted text within the discovered mementos.
\end{itemize}

\begin{figure}[!t]
\centering
\begin{Verbatim}[commandchars=\\\{\}]
\{
  \userinput{"updated"}: "2014-12-03T04:47:21.863568+00:00",
  "group": "__world__",
  "target": [
      \{
          "scope": ["http://climatefeedback.org"],
          "selector": [\{
                  "endContainer": "/div[2]/p[1]",
                  "endOffset": 57,
                  "type": "RangeSelector",
                  "startOffset": 0,
                  "startContainer": "/div[2]/p[1]"
              \},\{
                  "start": 50,
                  "end": 107,
                  "type": "TextPositionSelector"
              \},\{
                  \userinput{"exact"}: "Scientific feedback for Climate
                            Changeinformation online",
                  "prefix": "For Scientists CLIMATE FEEDBACK",
                  \userinput{"type"}: "TextQuoteSelector",
                  "suffix": "ABOUT feedback (noun): \"Helpful"
              \}
          ],
          "pos": \{"top": 148,"height": 25\},
          \userinput{"source"}: "http://climatefeedback.org/"
      \}
  ],
  "tags": [],
  \userinput{"text"}: "After reading about your project at MIT news, 
           I visited your page and got thoroughly confused
           since I didn't see a basic 'how to.' After 
           fumbling with the plugin, I didn't see the
           'channel' referred to on the page.  Finally,
           I saw [this post] (http://oceans.mit.edu/
           featured-stories/ climate-feedback)which 
           clarified that "official article evaluations
           will begin in the spring of 2015."I suggest
           adding a note about this to your site for
           wayward discoverers like myself.And good 
           luck! I'll be a the AGU and look forward
           to meeting you. ",
  "created": "2014-12-03T04:46:57.630434+00:00",
  "uri": "http://climatefeedback.org/",
  "user": "acct:branto@hypothes.is",
  "document": \{\userinput{ ... }\},
  "consumer": "00000000-0000-0000-0000-000000000000",
  "id": "xNec2gjYT5-ORcDg4fl7nA",
  "permissions": \{
      "admin": ["acct:branto@hypothes.is"],
      "read": ["acct:branto@hypothes.is","group:__world__"],
      "update": ["acct:branto@hypothes.is"],
      "delete": ["acct:branto@hypothes.is"]
  \}
\}
\end{Verbatim}
\caption{An Annotation Described in JSON Format, Available at https://hypothes.is/api/annotations/xNec2gjYT5-ORcDg4fl7nA}
\label{img:json}
\end{figure}

In Table \ref{tab:top_hosts}, we show the top 10 hosts with annotations at Hypothes.is.  Many of these hosts, including the top three, are academic servers and appear to use the system for annotation of scholarly work. Apart from this listing, we did not attempt to make judgements about the content of the annotations or annotation target text in our analysis. 

\begin{table}[!t]
\centering
\begin{tabular}{ r @{\hskip 0.2cm} | @{\hskip 0.2cm} l} 
\hline
\textbf{\shortstack{Number of Annotations}} & \textbf{\shortstack{Host}}\\ \hline
 1222 & \url{ caseyboyle.net}\\
 1191 & \url{ www.perseus.tufts.edu}\\
  887 & \url{rhetoric.eserver.org}\\
  875 & \url{networkedlearningcollaborative.com}\\
  749 & \url{sosol.perseids.org}\\
  733 & \url{tkbr.ccsp.sfu.ca}\\
  526 & \url{shakespeare.mit.edu}\\  
  391 & \url{hypothes.is}\\
  356 & \url{renaissancejohnson.weebly.com}\\  
  336 & \url{moodle2.wesleyan.edu}\\
\hline\end{tabular}
\vspace{0.5em}
\caption {The Top Hosts with Annotated Pages}
\label{tab:top_hosts}
\end{table}
 
\subsection{Determining the HTTP Status}

In the first step, the current HTTP status of annotation target URIs can be obtained by issuing HTTP HEAD requests for all URIs. In addition, we extended this to detect ``soft" 401, 403, and 404 URIs, which return a 200 OK status but actually indicate that the page is not found or is located behind authentication \cite{988716}. One technique we used to detect ``soft'' 4xx is to modify the original URI by adding some random characters to the parent directory, so that it is likely that the new URI does not exist. After that, we download the content of the original URI and the new one. If the content of both web pages is 93\% (or above) identical, and the HTTP status code of these URIs is ``200 OK'', then we consider that the HTTP status of the original URI is a ``soft'' 4xx. We have written a Python program available in GitHub\footnote{\url{https://github.com/maturban/Soft_4xx}} for detecting ``soft'' 4xx URIs.

The returned responses will determine the next action which should be made for every URI. The resulting responses can be categorized into three different groups. The first group contains URIs with hostnames  \texttt{localhost} or URIs which are actually URNs. The second group has URIs with one of the following status codes: ``soft'' and actual 400, 401, 403, 404, 429 or Connection-Timeout. URIs with 200 status code belong to the third group.

The first group, localhost and URNs, were excluded completely from our analysis because these are pages that are not publicly accessible on the live Web.  URIs in the second group, soft/actual 4xx and timed-out URIs, have been checked for mementos in the web archives. For URIs with response code 200, we have compared their associated highlighted annotation text with both the current version of the web page and the available mementos in the archives. Even though some annotations are still attached to their live web pages, we are interested to see if they have mementos to know how likely those annotations are to become orphans if their current web pages change or become unavailable.

\subsection{Are Annotations Attached to the Live Web?}
\label{subsec:checkLiveWeb}

The second step is to compare the annotated text (``\texttt{exact}'') of each annotation target URI that has a 200 HTTP status code with the current version of its web page and see if they match; this can be done by downloading the web page and extracting only the text which will be compared to the highlighted annotation text. Extracting text from a standard HTML resource is different from doing it with a Portable Document Format (PDF) file. We use \texttt{curl} to access and download web pages. Then, for HTML web resources, we extract only the text after cleaning it by removing all HTML tags, extra white-space characters, and others. If the web resource is a PDF file, extracting text will need one more step of converting the original binary contents to plain text. \texttt{pdftotext}\footnote{http://www.foolabs.com/xpdf/download.html}, which runs from the command-line and is freely available in many Linux distributions, can extract plain text from PDF files.  After extracting the text either from a standard HTML web page or a PDF file, we search for the highlighted annotation text. If the text is not found, the annotation is considered \emph{not attached}. For example, as shown in Figure \ref{img:live-page}, the annotation text is no longer attached to the web page as the highlighted text, shown in Figure \ref{img:arch-page}, has been removed from the live web page. For all annotations that are not attached until this point, we use PhantomJS to download the web page again trying to capture more resources that are loaded by JavaScript \cite{brunelle2013impact,BrunelleIpres15}.  

\begin{figure}[h!]
\centering
\includegraphics[width=90mm]{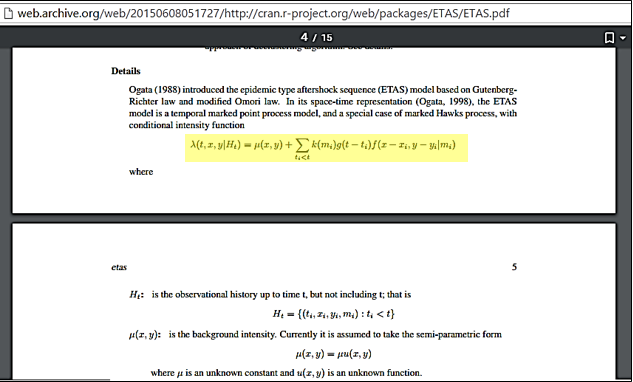}
\caption{Example of an Annotation That Can Not Be Re-attached to a PDF File (http://web.archive.org/web/20150608051727/http://cran.r-project.org/web/packages/etas/etas.pdf)}
\label{img:pdfexample}
\end{figure}

\begin{figure}[h!]
\centering
\includegraphics[width=90mm]{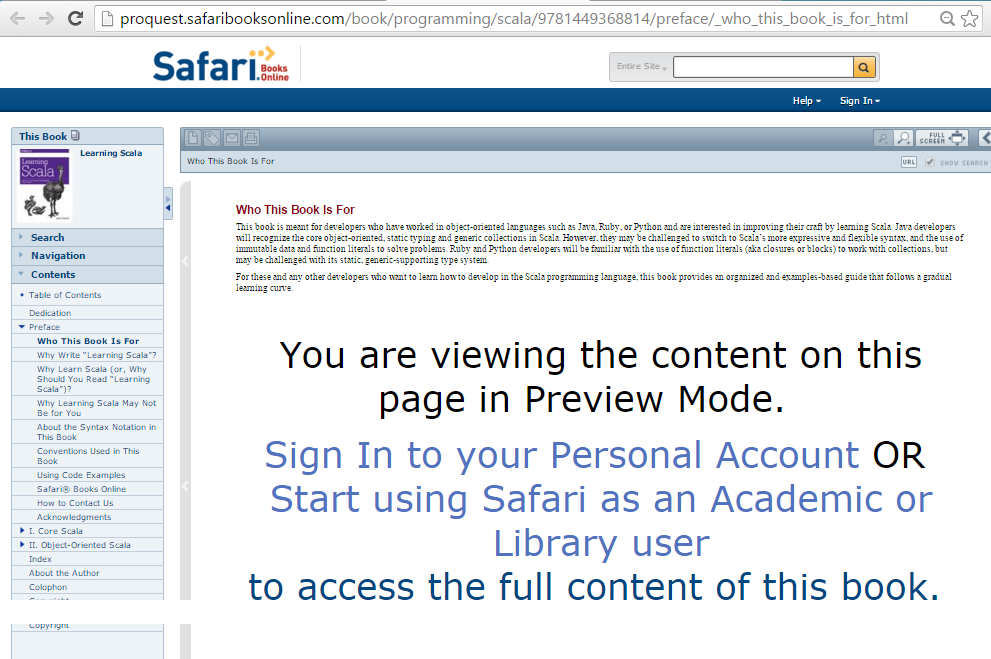}
\caption{Example of a Web Page (http://proquest.Safaribooksonline.Com/\\book/programming/scala/9781449368814/preface/\_who\_this\_book\_is\_for\_html) That Needs Authentication }
\label{img:soft}
\end{figure}

It is important to mention that when examining the existence of the annotated text in the related web pages, there are false negatives. In other words, our program might not identify some truly attached annotations because of reasons including but not limited to, not detecting ``soft'' 4xx URIs, failure in extracting text from PDF files by \texttt{pdftotext}, missing embedded resources in web pages that are loaded by JavaScript even with the use of PhantomJS. For example, the annotation that is about the highlighted equation in Figure \ref{img:pdfexample} can not be re-attached to the the PDF file\footnote{\url{http://web.archive.org/web/20150608051727/http://cran.r-project.org/web/packages/ETAS/ETAS.pdf}}  because the highlighted equation is saved with a different encoding. Figure \ref{img:soft} illustrates another example where a web page\footnote{\url{http://proquest.safaribooksonline.com/book/programming/scala/9781449368814/preface/_who_this_book_is_for_html}} needs authentication to access the missing content. We can consider this example as an undetected  ``soft'' 4xx as the server in this case should respond with 401 HTTP status code, but it returns ``200 OK'' instead.

\subsection{Discovering Mementos for All Valid URIs}
\label{subsubsec:timeGate}
The third step is to discover mementos for all valid annotation target URIs. For this purpose, we used a Memento Aggregator \cite{memento:rfc,nelson:memento:tr} which provides a TimeGate through which we can get mementos that are closest to a URI-R's date. It would be a time-consuming task to check all available mementos for a URI-R to see whether they can be used to recover web pages. For example, URIs like \url{http://www.nytimes.com/} or \url{http://www.cnn.com/} have thousands of mementos in different archives. The strategy that we use here is efficient in terms of execution time. For each URI, we only retrieve the nearest mementos to the annotation's creation date (``\texttt{updated}''). More precisely, we are capturing the closest memento to the date \textit{before} the annotation was created and the closest memento to the date \textit{after} the annotation was created.

\begin{figure}[ht]
\centering
\subfigure[Existing Mementos Before and After the Annotation Creation Date]{
\includegraphics[width=3.7in]{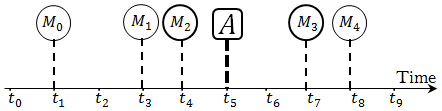}
\label{img:lr}
}\\
\subfigure[Existing Mementos Only Before the Annotation Creation Date]{
\includegraphics[width=3.7in]{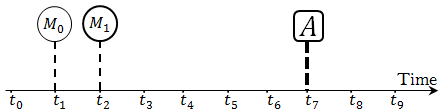}
\label{img:l}
}
\subfigure[Existing Mementos Only After the Annotation Creation Date]{
\includegraphics[width=3.7in]{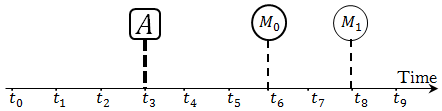}
\label{img:r}
}
\subfigure[No Mementos for the Annotation Target]{
\includegraphics[width=3.7in]{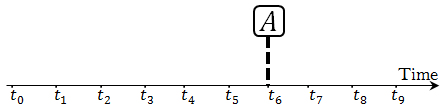}
\label{img:no}
}
\caption{Annotation and Memento Creation Dates}
\label{TS}
\end{figure}

In Figure \ref{img:lr}, the annotation \textit{A} was created at the time \textit{t\textsubscript{5}}. The closest memento to the date before \textit{t\textsubscript{5}} was \textit{M\textsubscript{2}} (captured at \textit{t\textsubscript{4}}) while the closest memento to the date after \textit{t\textsubscript{5}} was \textit{M\textsubscript{3}} (captured at \textit{t\textsubscript{7}}). So, for this annotation we picked the two closest mementos, which are \textit{M\textsubscript{2}} and \textit{M\textsubscript{3}}. Figure \ref{img:l} is an example where mementos are only available before the annotation creation date while in Figure \ref{img:r}, mementos are only available after the annotation creation date. It is also possible that an annotation target has no mementos at all, as Figure \ref{img:no} shows. If there are multiple closest mementos from different archives that share the same creation date (\textit{memento-datetime}), then we consider all of these mementos for two different reasons. First, it is possible that at the time a memento is requested from an archive, there would be a technical problem or server-related issue which may affect returning the requested mementos. Second, we would like to know how different archives could contribute to provide mementos and recover annotation target text.

\subsection{Are Annotations Attached to the Selected Mementos?}

The final step is to see whether annotated URIs can be recovered by their mementos. The same technique introduced in Section \ref{subsec:checkLiveWeb} is used to test mementos. If the annotation target text (``\texttt{exact}'') matches the text in the discovered memento, then we consider that this annotation is attached to the memento. Otherwise, we consider that the annotation cannot be attached.

\section{Results}

\begin{figure}[h!]
\centering
\includegraphics[width=105mm]{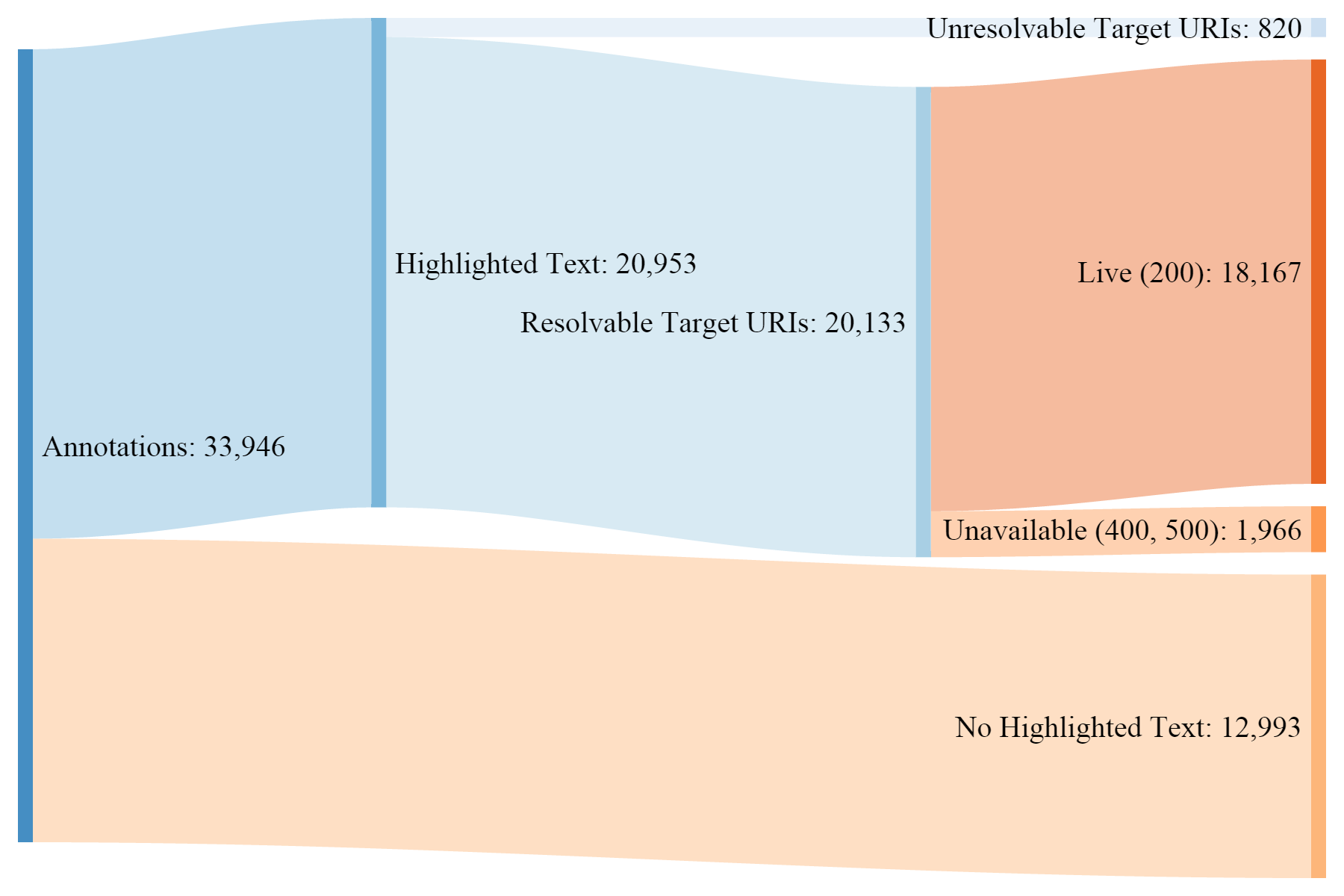}
\caption{HTTP Status Code of Annotations' Target URIs}
\label{img:status-code}
\end{figure}

We collected 33,946 annotations from Hypothes.is. Table~\ref{tab:uri-status} shows the results of checking the HTTP status code for the target URIs of all 20,953 highlighted text annotations. We noticed that a group of 820 annotations should be excluded from our analysis. This group consists of annotations whose target URIs are unresolvable, such as \texttt{localhosts} and URNs. In our further analysis, we will focus only on the 20,133 annotations that contain highlighted text and have resolvable target URIs. We noticed also that out of 20,133 annotations, 10\% (1966) have URI-Rs that are no longer available on the live web, returning 400 and 500 status codes. Figure \ref{img:status-code} shows how annotations are classified based on the status codes of their target URIs.

\begin{table}[!t]
\centering
\begin{tabular}{ r @{\hskip 0.2cm} | @{\hskip 0.2cm} l} 
\hline
\textbf{\shortstack{Number of Annotations}} & \textbf{\shortstack{Status Code}}\\ \hline
18167 & 200\\
778 & Time out\\
666 & 404\\
326 & \url{localhost}\\
318 & URN\\
190 & Soft 401/403/404\\
176	& Unknown\\
 87 & 401\\
 80 & 503\\   
 68	& 403\\  
 48 & 410\\   	  
 21 & 406\\
 19 & 500\\  
 5 & 416	\\
 2 & 520\\
 1 & 400	\\
 1 & 504\\
\hline\end{tabular}
\vspace{0.5em}
\caption {HTTP Status Code for All Annotation Target URIs}
\label{tab:uri-status}
\end{table}

\begin{figure}[h!]
\centering
\includegraphics[width=105mm]{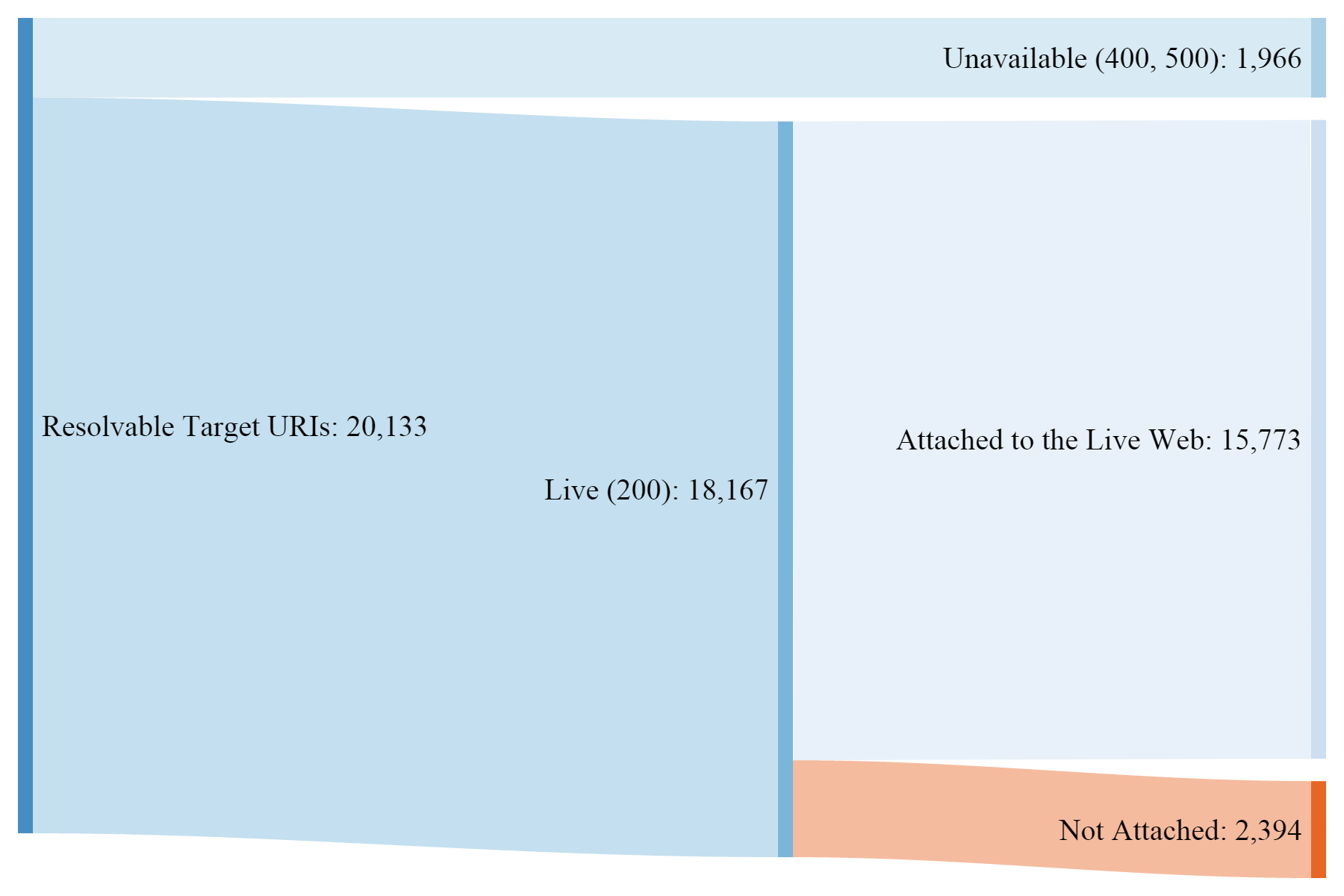}
\caption{Annotations that are Attached/Not Attached to the Live Web}
\label{img:attachnotattach}
\end{figure}

After checking each annotation, we found that 15,773 (78\%) of the highlighted text annotations are still attached to their live web pages as Figure \ref{img:attachnotattach} shows. This means that the remaining 22\% of the annotations are orphans if there are no mementos that can be used to reattach these annotations. 

Next for each annotation, we checked the archives for the presence of mementos of the target URI near the annotation creation date. In Table~\ref{tab:mem-left-right} we consider annotations that have mementos both before (``L'') and after (``R'') the annotation date. ``No'' under the L and R columns means that annotation cannot be attached to the nearest memento, while ``Yes'' means that the annotation attaches to the nearest memento.

\begin{table}
\centering
\begin{tabular}{r @{\hskip 0.2cm} |  @{\hskip 0.2cm} c @{\hskip 0.5cm} c @{\hskip 0.5cm} c}
\hline
\textbf{\shortstack{Number of\\Annotations}} & \textbf{\shortstack{Attached to\\Live Web Page}} & \textbf{\shortstack{ Attached  to\\Memento (L)}} & \textbf{\shortstack{Attached to\\Memento (R)}}\\
\hline
4091 & Yes & Yes & Yes\\
93 & Yes & Yes & No\\
100 & Yes & No & Yes\\
182 & Yes & No & No\\
251 & No & Yes & Yes\\
69 & No & Yes & No\\
44 & No & No & Yes\\
156 & No & No & No\\
\hline
\end{tabular}
\vspace{0.5em}
\caption {Annotation Targets with Existing Mementos Before and After the Annotation Creation Date.}
\label{tab:mem-left-right}
\end{table}

\begin{figure}[h!]
\centering
\includegraphics[width=105mm]{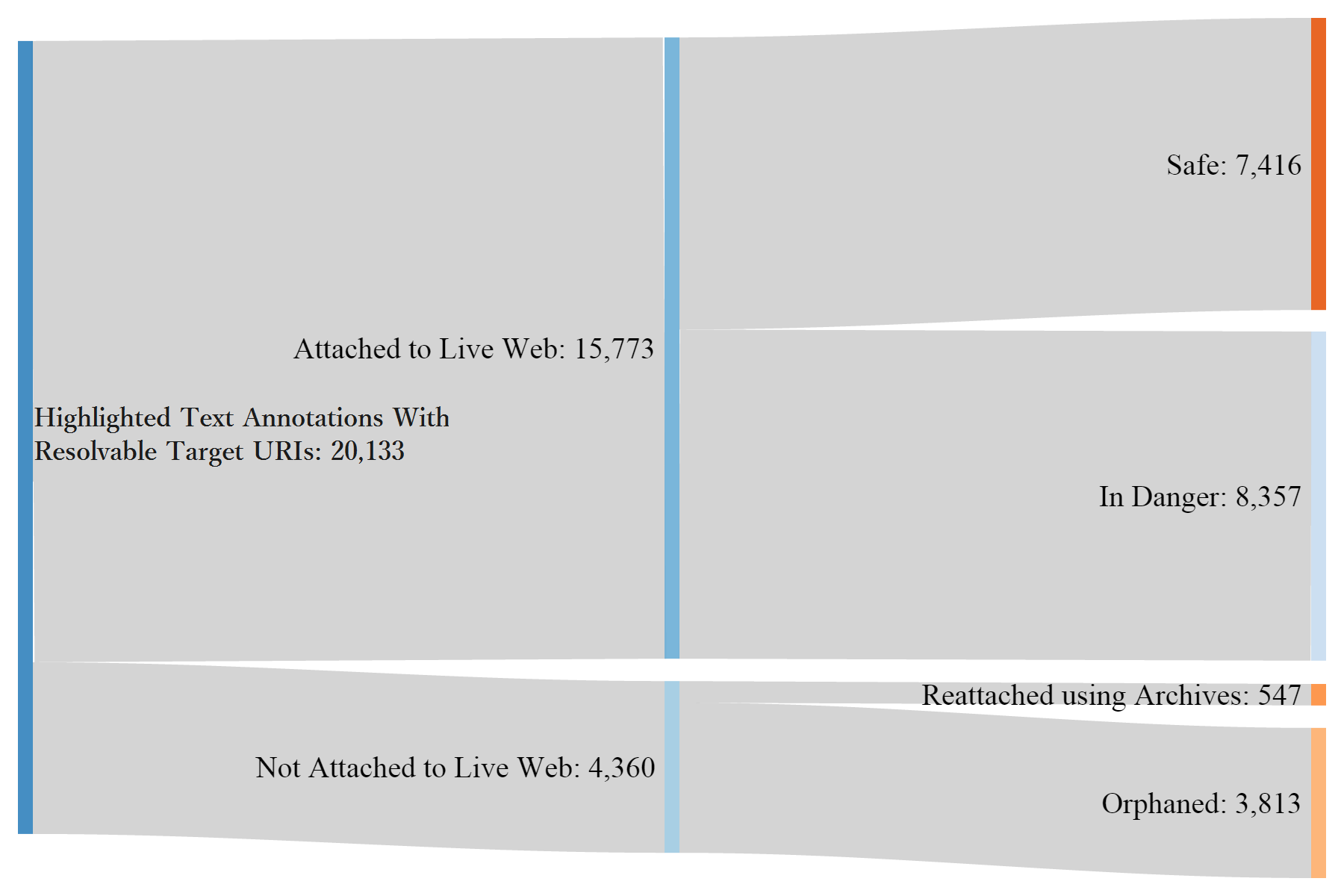}
\caption{The Status of Current Hypothes.is Annotations}
\label{img:overallstatus}
\end{figure}

Table~\ref{tab:mem-l-only} shows the number of annotations that have mementos only on the L side (before) of the annotation date, and Table~\ref{tab:mem-r-only} shows the number of annotations that have mementos only on the R side (after) of the annotation date. Finally, Table \ref{tab:mem-none} illustrates the number of annotations whose targets have no mementos. From these tables, we see that 4360 (22\%) of the annotations can no longer be attached to their live web pages. Unfortunately, the current holdings of web archives only allow 12\% of these to be re-attached, while the remaining 88\% of these annotations are orphans (19\% of all annotations included in our analysis). As shown in Table \ref{tab:mem-none}, the majority of annotations have no mementos available at all.  Those that can no longer be attached to their live web version are lost (orphans), but those that are still attached are in danger of being orphaned (41\% of all annotations). These annotations can be recovered if these pages are archived before the annotated text changes. As we can see, 60\% of annotations are either orphaned or in danger of being orphaned. Figure \ref{img:overallstatus} shows the status of current Hypothes.is annotations.  

Table~\ref{tab:archiveDist} shows the number of annotations that can be recovered using various archives, split by whether or not they are still attached to the live web. As expected the Internet Archive can be used to recover the most annotations.

\begin{table}[h]
\centering
\begin{tabular}{r @{\hskip 0.2cm} |  @{\hskip 0.2cm} c @{\hskip 0.5cm} c }
\hline
\textbf{\shortstack{Number of \\ Annotations}} & \textbf{\shortstack{Attached to Live\\ Web Page}}&\textbf{\shortstack{ Attached to \\ Memento (L)}}\\ \hline
1984&Yes&Yes\\
235&Yes&No\\
133&No&Yes\\
125&No&No\\ 
\hline\end{tabular}
\vspace{0.5em}
\caption {Annotation Targets with Existing Mementos Only Before the Annotation Creation Date}
\label{tab:mem-l-only}
\end{table}
\begin{table}[h]
\centering
\begin{tabular}{r @{\hskip 0.2cm} |  @{\hskip 0.2cm} c @{\hskip 0.5cm} c }
\hline
\textbf{\shortstack{Number of \\ Annotations}} & \textbf{\shortstack{Attached to Live\\ Web Page}}&\textbf{\shortstack{ Attached to \\ Mementos (R)}}\\ \hline
1397&Yes&Yes\\
101&Yes&No\\
50&No&Yes\\
98&No&No\\
\hline\end{tabular}
\vspace{0.5em}
\caption {Annotation Targets with Existing Mementos Only After the Annotation Creation Date}
\label{tab:mem-r-only}
\end{table}
\begin{table}[h]
\centering
\begin{tabular}{r @{\hskip 0.2cm} |  @{\hskip 0.2cm} c}
\hline
\textbf{\shortstack{Number of Annotations}} & \textbf{\shortstack{Attached to Live Web}}\\ \hline
7839&Yes\\
3434&No\\ 
\hline\end{tabular}
\vspace{0.5em}
\caption {Annotation Targets with No Existing Mementos}
\label{tab:mem-none}
\end{table}

\bgroup
\def\arraystretch{1.2}
\begin{table}[!t]
\centering
\begin{tabular}{l @{\hskip 0.2cm} |  @{\hskip 0.2cm} r @{\hskip 0.5cm} r }
\hline
\textbf{\shortstack{Archive}} &\textbf{\shortstack{Attached to \\Live Web}}&\textbf{\shortstack{Not Attached \\to Live Web}}\\ \hline
Internet Archive&6997 (94.30\%)&455 (83.10\%)\\
archive.is&679 (9.15\%)&39 (7.12\%)\\
Archive-It&562 (7.57\%)&47 (8.59\%)\\
github.com&80 (1.07\%)&21 (3.83\%)\\
wayback.vefsafn.is&71 (0.95\%)&53 (9.68\%)\\
arxiv.org&18 (0.24\%)&0\\
www.webarchive.org.uk&4 (0.05\%)&0\\
webarchive.loc.gov&3 (0.04\%)&0\\
webarchive.nationalarchives.gov.uk&2 (0.02\%)&0\\
discordia.wikia.com&1 (0.01\%)&0\\ 
Total&8417 (113.40\%)&615 (112.32\%)\\
\hline\end{tabular}
\vspace{0.5em}
\caption {Annotation Targets Recovered by Different Archives}
\label{tab:archiveDist}

\end{table}

\section{Conclusions}
In this paper, we analyzed the attachment of highlighted text annotations in Hypothes.is.  We studied the prevalence of orphaned annotations, and found that 19\% (3813) of the highlighted text annotations are orphans while 41\% (8357) are in danger of being orphaned. We used Memento to look for archived versions of the annotated pages and found that 3\% (547) of annotations that are not attached to the live web can be reattached to archived versions.  We also found that for the majority of the annotations (11,273), no memento exists in the archives. This points to the need for archiving web pages at the time of annotation. 

\bibliographystyle{splncs03}
\bibliography{annotations}

\end{document}